\documentclass[cits]{PoS}
\usepackage{amsmath}

\title{Lattice effective field theory for nuclei from A = 4 to A = 28}
\ShortTitle{Lattice effective field theory for nuclei from A = 4 to A = 28}

\author{
\speaker{Timo A. L\"ahde},$^a$ Evgeny Epelbaum,$^b$ Hermann Krebs,$^b$
Dean Lee,$^c$ Ulf-G.~Mei{\ss }ner,$^{ade}$ and Gautam~Rupak$^f$ \\
\llap{$^a$}Institute~for~Advanced~Simulation, Institut~f\"{u}r~Kernphysik, and 
J\"{u}lich~Center~for~Hadron~Physics,~Forschungszentrum~J\"{u}lich, D--52425~J\"{u}lich, Germany \\
\llap{$^b$}Institut~f\"{u}r~Theoretische~Physik~II,~Ruhr-Universit\"{a}t~Bochum, D--44870~Bochum,~Germany \\
\llap{$^c$}Department~of~Physics, North~Carolina~State~University, Raleigh, NC~27695, USA \\
\llap{$^d$}Helmholtz-Institut f\"ur Strahlen- und Kernphysik and Bethe Center for Theoretical Physics,
Universit\"at Bonn, D--53115 Bonn, Germany \\
\llap{$^e$}JARA~--~High~Performance~Computing, Forschungszentrum~J\"{u}lich, D--52425 J\"{u}lich,~Germany \\
\llap{$^f$}Department~of~Physics~and~Astronomy, Mississippi~State~University, Mississippi State, MS~39762, USA \\
E-mail: \email{t.laehde@fz-juelich.de}, \email{evgeny.epelbaum@rub.de}, \email{hermann.krebs@rub.de},
\email{dean\_lee@ncsu.edu}, \email{u.meissner@fz-juelich.de}, \email{grupak@u.washington.edu}
}

\abstract{
We present an overview of the extension of Nuclear Lattice Effective Field Theory simulations to the regime of medium-mass nuclei. We focus on
the determination of the ground-state energies of the alpha nuclei $^{16}$O, $^{20}$Ne, $^{24}$Mg and $^{28}$Si by means of Euclidean
time projection.} 


\FullConference{31st International Symposium on Lattice Field Theory LATTICE 2013 \\
		 July 29 -- August 3, 2013 \\
		 Mainz, Germany}

\begin{document}


\section{Introduction}

Nuclear Lattice Effective Field Theory (NLEFT) is a first-principles approach, in which Chiral EFT for nucleons is combined with 
numerical Auxiliary-Field Quantum Monte Carlo (AFQMC) lattice simulations. NLEFT differs from other \textit{ab initio} 
methods~\cite{Hagen:2012fb,Jurgenson:2013yya,Roth:2011ar,Hergert:2012nb,Lovato:2013cua,Duguet} in that it is an 
unconstrained Monte Carlo calculation, which does not rely on truncated basis expansions or many-body perturbation theory, nor on prior 
information about the structure of the nuclear wave function.


\section{Nuclear Lattice EFT at NNLO}

As in Chiral EFT, our calculations are organized in powers of a generic soft scale $Q$ associated with factors of momenta and the 
pion mass~\cite{Epelbaum:2008ga}. 
We denote $\mathcal{O}(Q^0)$ as leading order (LO), $\mathcal{O}(Q^2)$ as next-to-leading order (NLO), and $\mathcal{O}(Q^3)$
as next-to-next-to-leading order (NNLO) contributions. The present calculations are performed up to NNLO.
We define $H_\mathrm{LO}^{}$ as the LO lattice Hamiltonian, and $H_\mathrm{SU(4)}^{}$ as the equivalent Hamiltonian with
the pion-nucleon coupling $g_A^{} = 0$ and contact interactions that respect Wigner's SU(4) symmetry. 

In our NLEFT calculations (see Ref.~\cite{Dean_QMC} for a review), $H_\mathrm{LO}^{}$ is treated non-perturbatively.
The NLO contribution to the two-nucleon force (2NF), the electromagnetic and strong isospin-breaking contributions (EMIB), 
and the three-nucleon force (3NF) which first enters at NNLO, are all treated as perturbations. It should be noted that our ``LO'' calculations 
use smeared short-range interactions that capture much of the corrections usually treated at NLO~\cite{Borasoy:2006qn}.
The 3NF at NNLO over-binds 
nuclei with $A > 4$ due to a clustering instability which involves four nucleons on the same lattice site. The long-term objective of NLEFT is to
remedy this problem by decreasing the lattice spacing and including the N3LO corrections in Chiral EFT. In the mean time, the over-binding problem
has been rectified by means of a 4N contact interaction, tuned to the empirical binding energy of either $^4$He or 
$^8$Be~\cite{Epelbaum:2009pd}. While this provides a good description of the alpha nuclei up to $A = 12$ including the Hoyle 
state~\cite{Epelbaum:2009pd,Epelbaum:2011md,Epelbaum:2012qn}, the 
over-binding is found to increase more rapidly for $A \geq 16$. 
Therefore, in Ref.~\cite{A28_letter} a non-local 4N interaction which accounts for all possible configurations of four nucleons on adjacent lattice sites was
introduced, and adjusted to the empirical binding energy of $^{24}$Mg.
A detailed study of the spectrum of $^{16}$O will be reported separately~\cite{16O_spectrum}.


\section{Euclidean time projection}

The NLEFT calculations reported here (see also Ref.~\cite{A28_letter}) are performed with a (spatial) lattice spacing of $a=1.97$~fm
in a periodic cube of length $L = 11.8$~fm. Our trial wave function
$|\Psi_{A}^\mathrm{init}\rangle$ is a Slater-determinant state composed of delocalized standing waves,
with $A$ nucleons and the desired spin and isospin. First, we 
project $|\Psi_{A}^\mathrm{init}\rangle$ for a time $t^\prime$ using the Euclidean-time evolution operator of the 
SU(4) Hamiltonian, giving the ``trial state'' 
$|\Psi_A^{}(t^\prime_{})\rangle \equiv \exp(-H_\mathrm{SU(4)}^{} t^\prime_{}) |\Psi_{A}^\mathrm{init}\rangle$.
Second, we use the full Hamiltonian $H_\mathrm{LO}^{}$ to construct the Euclidean-time projection amplitude
\begin{equation}
Z_A^{}(t) \equiv \langle\Psi_A^{}(t^\prime_{})| \exp(-H_\mathrm{LO}^{} t) |\Psi_A^{}(t^\prime_{})\rangle,
\qquad
E_A^{}(t) = -\partial[\ln Z_A^{}(t)]/\partial t,
\label{EAt}
\end{equation}
and the ``transient energy'' $E_A^{}(t)$. If we denote
by $|\Psi_{A,0}^{}\rangle$ the lowest (normalizable) eigenstate of $H_\mathrm{LO}^{}$
which has a non-vanishing overlap with the trial state $|\Psi_A^{}(t^\prime_{})\rangle$,
we obtain the corresponding energy $E_{A,0}^{}$ as the ${t\to\infty}$ limit of $E_A^{}(t)$.

\begin{figure}[t!]
\begin{center}
\includegraphics[width=.92\columnwidth]{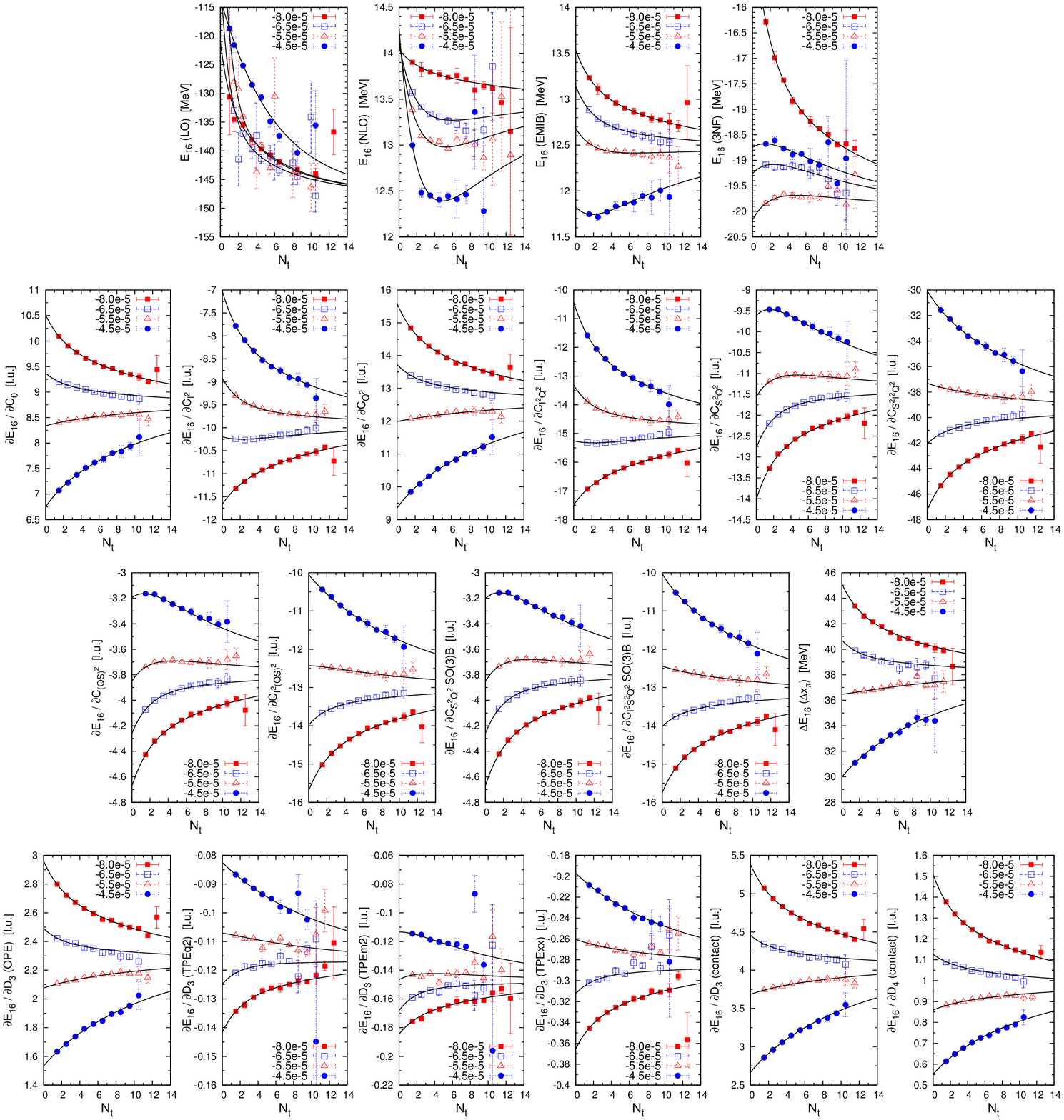}
\vspace{-.7cm}
\end{center}
\caption{NLEFT results for $^{16}$O. The LO energy is $E_\mathrm{LO}^{} = -147.3(5)$~MeV, and the result
at NNLO including 4N interactions is $E_\mathrm{NNLO+4N}^{} = -131.3(5)$~MeV. The empirical binding energy is
$-127.62$~MeV.
\label{16O}}
\end{figure}

The NLO and NNLO contributions are evaluated in perturbation theory. We compute
operator expectation values using
\begin{equation}
Z_A^\mathcal{O}(t) \equiv \langle\Psi_A^{}(t^\prime_{})| \exp(-H_\mathrm{LO}^{} t/2)
\mathcal{O} \exp(-H_\mathrm{LO}^{} t/2)  |\Psi_A^{}(t^\prime_{})\rangle,
\label{OP}
\end{equation}
for any operator $\mathcal{O}$. Given the ratio $X_A^\mathcal{O}(t) = Z_A^\mathcal{O}(t)/Z_A^{}(t)$, the expectation value 
of $\mathcal{O}$ for the desired state $|\Psi_{A,0}^{}\rangle$ is obtained as 
$X_{A,0}^\mathcal{O} \equiv \langle\Psi_{A,0}^{}| \mathcal{O} |\Psi_{A,0}^{}\rangle = \lim_{t \to \infty}X_A^\mathcal{O}(t)$.

\begin{figure}[t!]
\begin{center}
\includegraphics[width=.92\columnwidth]{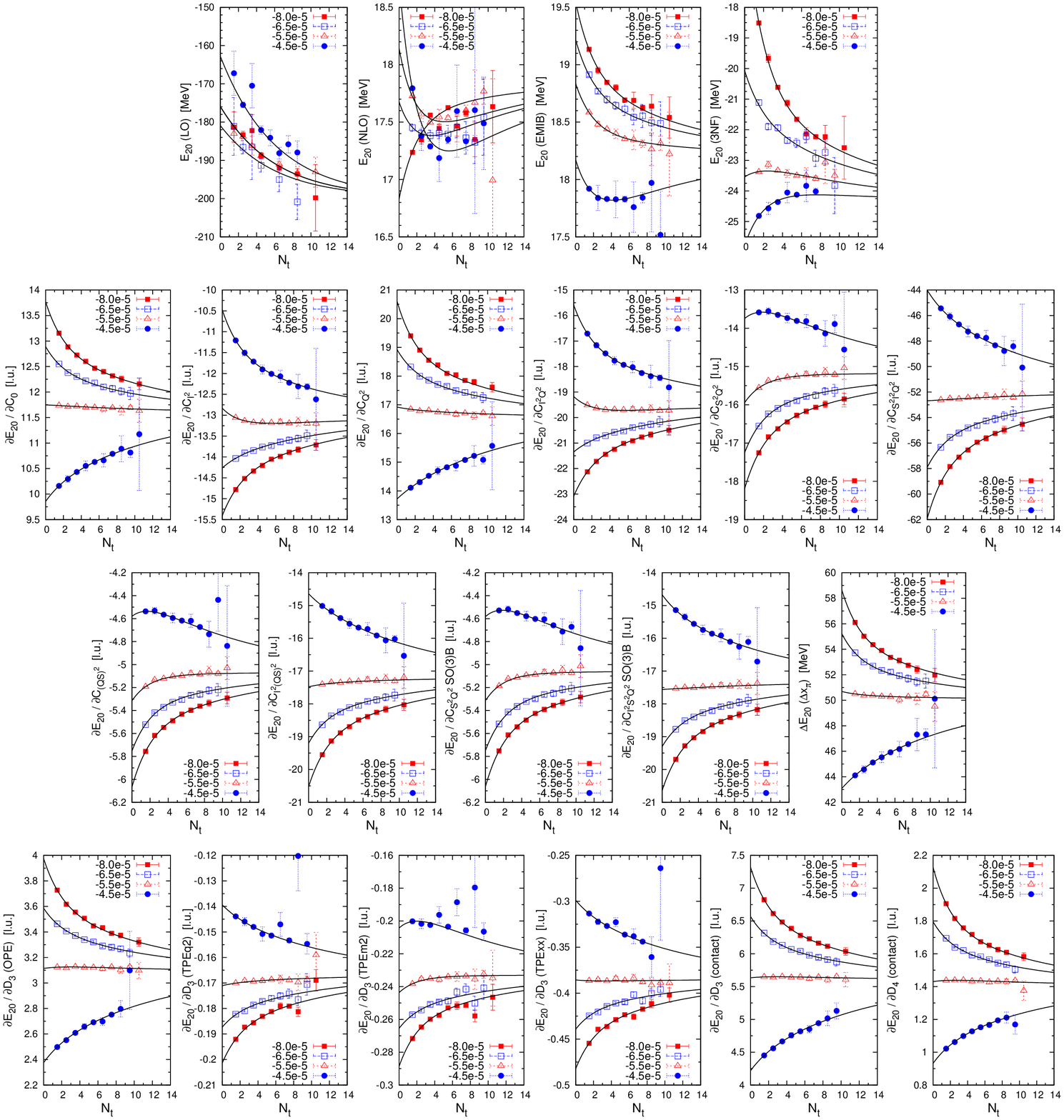}
\vspace{-.7cm}
\end{center}
\caption{NLEFT results for $^{20}$Ne. The LO energy is $E_\mathrm{LO}^{} = -199.7(9)$~MeV, and the result
at NNLO including 4N interactions is $E_\mathrm{NNLO+4N}^{} = -165.9(9)$~MeV. The empirical binding energy is
$-160.64$~MeV.
\label{20Ne}}
\end{figure}

\begin{figure}[t!]
\begin{center}
\includegraphics[width=.92\columnwidth]{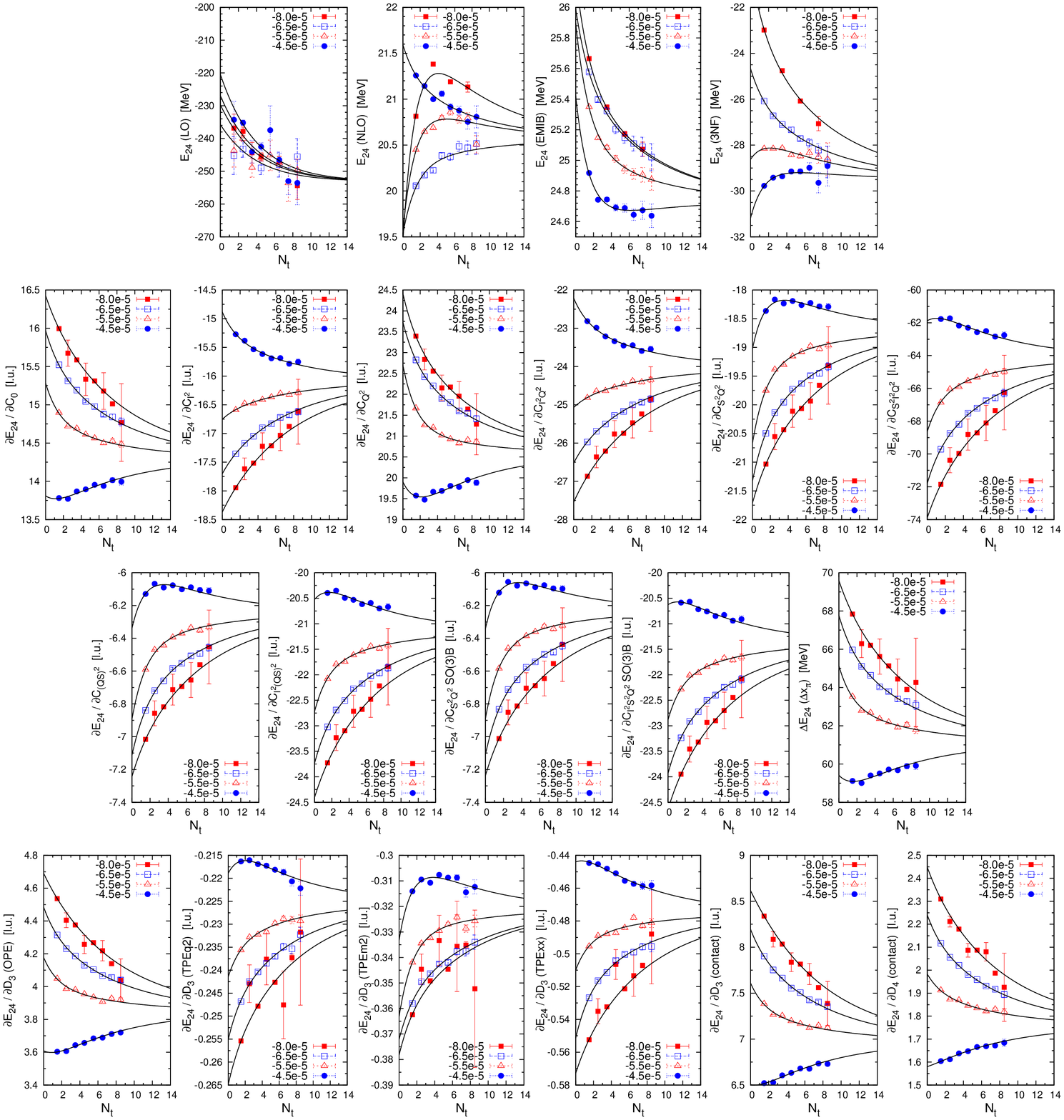}
\vspace{-.7cm}
\end{center}
\caption{NLEFT results for $^{24}$Mg. The LO energy is $E_\mathrm{LO}^{} = -253(2)$~MeV, and the result
at NNLO including 4N interactions is $E_\mathrm{NNLO+4N}^{} = -198(2)$~MeV. The empirical binding energy is
$-198.26$~MeV.
\label{24Mg}}
\end{figure}

\begin{figure}[t!]
\begin{center}
\includegraphics[width=.92\columnwidth]{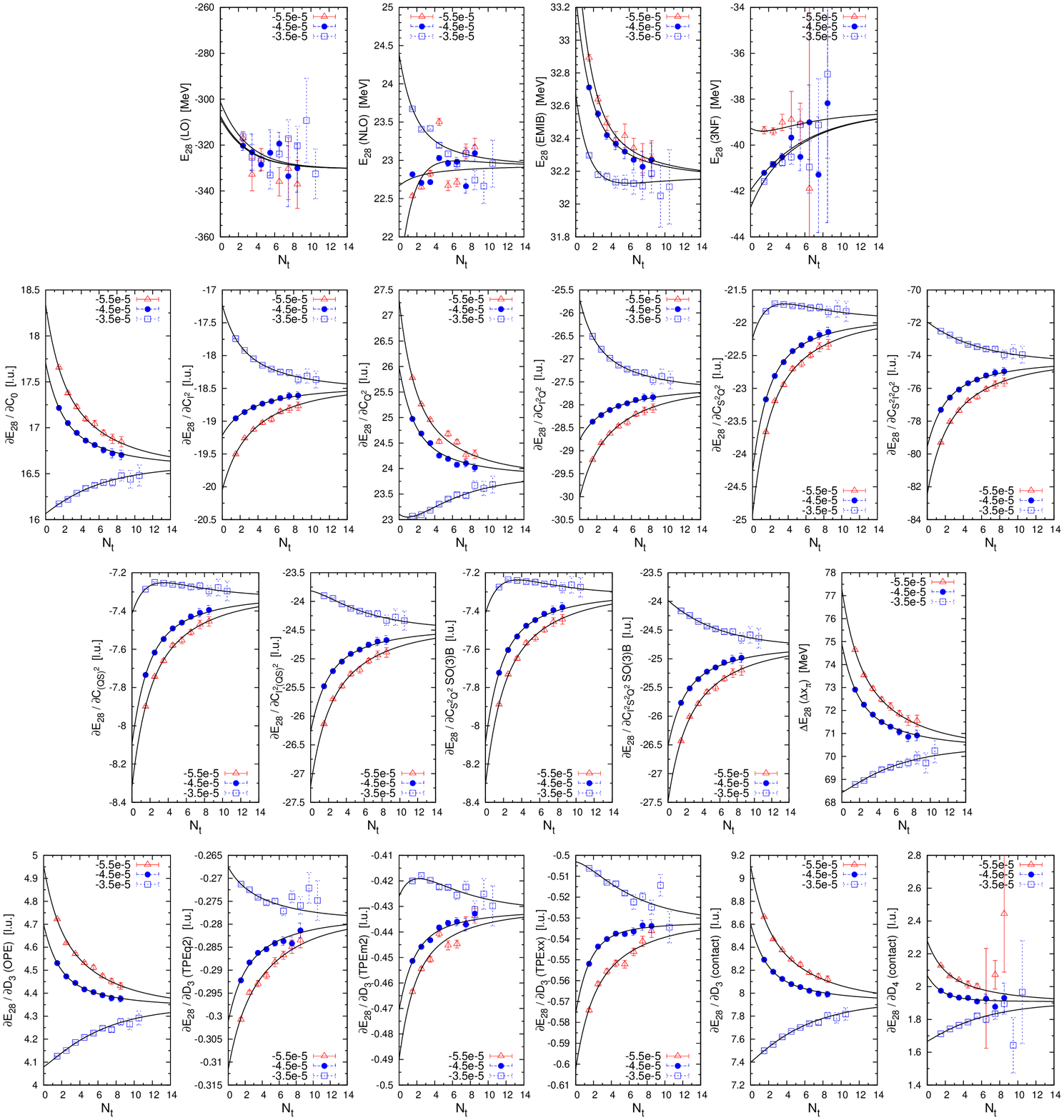}
\vspace{-.7cm}
\end{center}
\caption{NLEFT results for $^{28}$Si. The LO energy is $E_\mathrm{LO}^{} = -330(3)$~MeV, and the result
at NNLO including 4N interactions is $E_\mathrm{NNLO+4N}^{} = -233(3)$~MeV. The empirical binding energy is
$-236.54$~MeV.
\label{28Si}}
\end{figure}

Sign oscillations make it difficult to reach sufficiently large values of the 
projection time~$t$. It is helpful to note that the closer the trial state $|\Psi_A^{}(t^\prime_{})\rangle$ is to $|\Psi_{A,0}^{}\rangle$, 
the less the necessary projection time $t$. 
$|\Psi_A^{}(t^\prime_{})\rangle$ can be optimized by adjusting both the SU(4) projection time $t^\prime_{}$ and the strength 
of the coupling $C_\mathrm{SU(4)}^{}$ of $H_\mathrm{SU(4)}$. 
The accuracy of the extrapolation $t\to\infty$ can be further improved by simultaneously incorporating data from trial states that differ in 
$C_\mathrm{SU(4)}^{}$. 

The large-time behavior of $Z_A^{}(t)$ and $Z_A^{\mathcal{O}}(t)$ is controlled by the low-energy spectrum of $H_\mathrm{LO}^{}$. 
Let $| E\rangle$ label the eigenstates of $H_\mathrm{LO}^{}$ with energy $E$, and let $\rho_{A}^{}(E)$ denote the density of states for a system
of $A$~nucleons. We then express $Z_A^{}(t)$ and $Z_A^{\mathcal{O}}(t)$ in terms of their spectral representations,
\begin{align}
Z_A^{}(t) = & \int dE \: \rho_A^{}(E) \:
\big| \langle E |\Psi_A^{}(t^\prime_{})\rangle\big|^2_{} 
\exp(-Et), \\
Z_A^{\mathcal{O}}(t) = & \int dE\,dE^\prime_{} \, \rho_A^{}(E)\,\rho_A^{}(E^\prime_{}) 
\langle\Psi_A^{}(t^\prime_{})|E\rangle \,
\langle E|\mathcal{O}|E^\prime_{}\rangle \,
\langle E^\prime_{}|\Psi_A^{}(t^\prime_{})\rangle \, \exp(-(E+E^\prime_{})t/2),
\end{align} 
from which we construct the spectral representations of $E_A^{}(t)$ and $X_A^{\mathcal{O}}(t)$. 
We can approximate these to arbitrary accuracy over any finite range of $t$ by taking $\rho_{A}^{}(E)$ to be a sum of energy delta functions,
$\rho_{A}^{}(E) \approx \sum_{i=0}^{i_\mathrm{max}}\delta(E-E_{A,i}^{})$,
where we take $i_\mathrm{max} = 4$ for the $^4$He ground state, 
and $i_\mathrm{max} = 3$ for $A \geq 8$. Using data obtained for
different values of $C_\mathrm{SU(4)}^{}$, we perform a correlated fit of $E_A^{}(t)$ and $X_A^{\mathcal{O}}(t)$ for all operators $\mathcal{O}$
that contribute to the NLO and NNLO energy corrections. We find that the use of $2-6$ trial states allows for a much more precise
determination of $E_{A,0}^{}$ and $X_{A,0}^{\mathcal{O}}$ than hitherto possible. In particular, we may ``triangulate'' 
$X_{A,0}^{\mathcal{O}}$ using trial states that correspond to functions $X_A^{\mathcal{O}}(t)$ which converge both from above and below,
thereby bracketing $X_{A,0}^{\mathcal{O}}$.


\section{Results}

The NLEFT results for $^{16}$O are given in Fig.~\ref{16O}, for $^{20}$Ne in Fig.~\ref{20Ne}, 
for $^{24}$Mg in Fig.~\ref{24Mg}, and for $^{28}$Si in Fig.~\ref{28Si}. The curves show a correlated fit for all trial states, using the same 
spectral density $\rho_A^{}(E)$. The upper row in each figure shows the LO energy, the total isospin-symmetric 
2NF correction (NLO), the electromagnetic and isospin-breaking corrections (EMIB) and the total 3NF correction. 
The remaining panels show the matrix elements $X_A^{\mathcal{O}}(t)$ that form part of the NLO and 3NF terms. 
The operators $\partial E_{A}^{}/\partial C_i^{}$ give the contributions of the NLO contact interactions, and 
$\Delta E_{A}^{} (\Delta x_\pi^{})$ denotes the energy shift due the $\mathcal{O}(a^2)$-improved pion-nucleon coupling. 
The operators $\partial E_{A}^{}/\partial D_i^{}$ give the individual contributions to the total 3NF correction. 

To summarize, we have reported on the extension of NLEFT to the regime of medium-mass nuclei.
While the NNLO results are good up to $A = 12$, an increasing
over-binding (associated with the momentum-cutoff scale and neglected
higher-order contributions) manifests itself for $A \geq 16$. While the
long-term objectives of NLEFT are to decrease the lattice spacing and include higher orders in the EFT expansion, we also find that the missing 
physics can be approximated by an effective 4N interaction. 
The current exploratory results represent an important step 
towards more comprehensive NLEFT simulations of medium-mass nuclei in the future.

\acknowledgments

We are grateful for the help in automated data collection by Thomas Luu.
Partial financial support from the Deutsche Forschungsgemeinschaft (Sino-German CRC 110), the Helmholtz Association (Contract No.\ VH-VI-417), 
BMBF (Grant No.\ 06BN9006), and the U.S. Department of Energy (DE-FG02-03ER41260) is acknowledged. This work was further supported
by the EU HadronPhysics3 project, and funds provided by the ERC Project No.\ 259218 NUCLEAREFT. The computational resources 
were provided by the J\"{u}lich Supercomputing Centre at the Forschungszentrum J\"{u}lich and by RWTH Aachen.


\end{document}